\documentstyle[aps]{revtex}

\title{Quantum dynamics in canonical and micro-canonical ensembles.
Part II. Tunneling in double well potential.}

\author{
V.S. Filinov \thanks{Other author information: (Send correspondence to V.S.F.)
V.S.F.: Email: filinov@vovan.msk.ru; Telephone: 7(095)931-07-19;
Fax: 7(095)485-79-90}, \ \\
{\it Russian Academy of Sciences, 'IVTAN' Association \ \ \\
High Energy Density Research Center \ \ \\
Izhorskaya str. 13/19, Moscow,
\hspace{0.5em}127412, \hspace{0.5em}Russia} \\
Yu. E. Lozovik, A. V. Filinov \ \\
{\it Russian Academy of Sciences, Institute of Spectroscopy,  \ \ \\
\hspace{0.5em}Troitsk, Moscow region, 142092, \hspace{0.5em}Russia} \ \\
I. Zacharov \ \ \\
{\it Silicon Graphics Computer Systems, SGI Europe, \ \ \\
Grand Atrium, route des Avouillons 30, \\
\hspace{0.5em}1196  Gland, \hspace{0.5em}Switzerland}  \ \\
Alexei M. Oparin \\
{\it Russian Academy of Sciences, Institute for Computer Aided Design \ \\
Vtoraya Brestskaya str. 19/18, \ \\
\hspace{0.5em}Moscow, 123056, \hspace{0.5em}Russia}
}

\begin{document}
\maketitle
\hspace{3cm} Phys. Abstr. Class.: 72.10.Bg; 02.10.-c; 02.70-c; 02.50-r; 02.60-x
\begin{abstract}
In the second part of this paper in micro canonical ensemble
the new numerical approach for consideration of quantum dynamics and
calculations of the average values of quantum operators and time correlation
functions in the Wigner representation of quantum statistical mechanics has
been developed. The time correlation functions have been presented in the
form of the integral of the Weyl's symbol of considered operators and the
Fourier transform of the product of matrix elements of the dynamic
propagators. For the last function the integral Wigner- Liouville's type
equation has been derived. The initial condition for this equation has been
obtained in the form of the Fourier transform of the Wiener path integral
representation of the matrix elements of the propagators at initial time.
The numerical procedure for solving this equation combining both molecular
dynamics and Monte Carlo methods has been developed.

The numerical results have been obtained for series of the average values of
quantum operators as well as for the time correlation function
characterizing the energy level structure, the momentum flow of tunneling
particles at barrier crossing and the absorption spectra of electron in
potential well. The developed quantum dynamics
method was tested by comparison of numerical results with analytical
estimations. Tunneling transitions and the effect of the quasi stationary
state has been considered as the reason of the peculiarities in behaviour of
the time correlation functions and position and momentum dispersions.

Possibility of applying the developed approach to the theory of classical
wave propagation in random media have been also considered. For classical
waves some results have been obtained for Gaussian beam propagation in 2D
and 3D waveguides.

\end{abstract}

\date{July 1997}


\newpage




\section{Introduction}

In canonical ensemble considered in the first part of this paper \cite{can1}
the numerical studies of the exponentially small tunneling effects is very
difficult due to the temperature averaging. To overcome this difficulty the
more delicate approach in the micro- canonical ensemble with a fixed initial
energy has been introduced by the inverse Laplace transformation of the
spectral density on the inverse temperature variable. The related integral
Wigner- Liouville's type equation have been obtained and the numerical
approach combining both molecular dynamics and Monte Carlo methods for
solving this equation has been developed.

The time correlation function characterizing the energy level structure, the
momentum flow of tunneling particles at barrier crossing and the absorption
spectra of electron in potential well have been calculated.
Tunneling transitions and the effect of the quasi stationary
state has been considered as the reason of the peculiarities in behaviour of
the time correlation functions and position and momentum dispersions.

Possibility of applying the developed approach to the theory of classical
wave propagation in random media have been also considered. For classical
waves some results have been obtained for Gaussian beam propagation in 2D
and 3D waveguides.

\section{Electron in double well potential}

In the second part of this paper we've considered dynamics of quantum
electron in a deep symmetric (with respect zero of x-axis $q^x=0$) double
well potential:
\[
V_0\tilde{U}\left( \left| q\right| \right) =V_0\left\{ V_0^{\prime
}/V_0*\exp \left( \left| q\right| ^2/\sigma ^2\right) -\exp \left( \left|
q\right| ^2/\tilde{\sigma}^2\right) \right\}
\]
where $V_0^{\prime }\ll V_0$ and $\tilde{\sigma}>\sigma $ ($V_0^{\prime
}/V_0=0.12$).

To analyze the tunneling effects, absorption spectra and electron energy
levels we have considered the Fourier transform of the time correlation
functions characterizing the quantum particle momentum flow through the
barrier
\begin{equation}
\begin{array}{c}
k\left( \omega ,E\right) =\int_0^\infty \exp \left( -i\omega t-\epsilon
t\right) C_{F\eta }\left( t,E\right) dt= \\
\tilde{Z}^{-1}\int_0^\infty \exp \left( -i\omega t-\epsilon t\right)
Tr\left( \hat{F}\exp \left( i\hat{H}t/\hbar \right) \hat{\eta}\exp \left( -i%
\hat{H}t/\hbar \right) \delta \left( E-\hat{H}\right) \right) dt= \\
\tilde{Z}^{-1}\sum_{\nu ,\mu }\left\langle \Psi _\mu |\hat{F}|\Psi _\upsilon
\right\rangle \left\langle \Psi _\mu |\hat{\eta}|\Psi _\upsilon
\right\rangle \delta _\epsilon (\left( E_\mu -E_\upsilon \right) /\hbar
-\omega )\delta _\epsilon (E-E_\mu )
\end{array}
\label{mc1}
\end{equation}
where $\epsilon \rightarrow 0$, $\delta \left( E-\hat{H}\right) $ is the
initial density matrix , $\tilde{Z}=Tr\left( \delta \left( E-\hat{H}\right)
\right) $, $E_\mu $ $\Psi _\mu $ are eigenvalues and eigenfunctions of the
Hamiltonian of the system. Due to delta- functions the function $k(\omega ,E)
$ should have peaks in discrete part of spectrum on the frequencies $\omega $
equal to the different combinations $\left( E_\mu -E_\upsilon \right) /h$.
Here $C_{F\eta }\left( t,E\right) $ is defined in the Wigner representation
by expression:
\begin{equation}
\begin{array}{c}
C_{F\eta }\left( t,E\right) = \\
\frac 1{\left( 2\pi h\right) ^{2\upsilon }}\int \int dp_1dq_1dp_2dq_2\frac
12\left( F\left( p_1^x,q_1^x\right) \eta \left( q_2^x\right) +F\left(
p_2^x,q_2^x\right) \eta \left( q_1^x\right) \right) \times  \\
W\left( p_1,q_1;p_2,q_2;t;E\right)
\end{array}
\label{mc2}
\end{equation}
where Weyl's symbols of operators $\hat{F}$ and $\hat{\eta}$ are: $F\left(
p,q\right) =\frac 1{2m}\left[ p\delta \left( q\right) +\delta \left(
q\right) p\right] $, $\eta $ is step function that projects onto $q^x>0$
half- space.

Introduced in the first part of this paper \cite{can1} the spectral
densities in canonical $W\left( p_1,q_1;p_2,q_2;t;i\hbar \beta \right) $ and
micro-canonical $W\left( p_1,q_1;p_2,q_2;t;E\right) $ ensembles are
connected according to the (\ref{mc1}) by the Laplace transformation:
\begin{eqnarray*}
W\left( p_1,q_1;p_2,q_2;t;i\hbar \beta \right)  &=&\int dE\exp \left( -\beta
E\right) W\left( p_1,q_1;p_2,q_2;t;E\right) = \\
&&\int dE\exp \left( -\beta E\right) \frac 1{2\pi }\int_{-\infty }^\infty
d\omega \exp \left( i\omega t-\epsilon t\right) \exp \left( \frac{\beta
\hbar \omega }2\right) k\left( \omega ,E\right)
\end{eqnarray*}
where
\[
\begin{array}{c}
W\left( p_{1,}q_1;p_{2,}q_2;t;E\right) =\tilde{Z}^{-1}\int \int d\xi _1d\xi
_2\exp \left( i\frac{p_1\xi _1}\hbar \right) \exp \left( i\frac{p_2\xi _2}%
\hbar \right) \times  \\
\times \left\langle q_1+\frac{\xi _1}2\left| \exp \left( i\hat{H}t/\hbar
\right) \right| q_2-\frac{\xi _2}2\right\rangle \left\langle q_2+\frac{\xi _2%
}2\left| \exp \left( -i\hat{H}t/\hbar \right) \delta \left( E-\hat{H}\right)
\right| q_1-\frac{\xi _1}2\right\rangle
\end{array}
\]

We suppose that function $W\left( p_1,q_1;p_2,q_2;t;E\right) $ provides more
information about quantum effects and quantum dynamics of the system.

The functions $W\left( p_1,q_1;p_2,q_2;t;E\right) $ as can be easily proved
are the solutions of the same linear integral equation as have been obtained
in the first part of this paper \cite{can1}.
\begin{equation}
W(p_1,q_1;p_2,q_2;t;E)=\bar{W}(\bar{p}_0,\bar{q}_0;\tilde{p}_0,\tilde{q}%
_0;E)+ \\
\int_0^td\tau \int dsd\eta W(\bar{p}_\tau -s,\bar{q}_\tau ;\tilde{p}_\tau
-\eta ,\tilde{q}_\tau ;\tau ;E)\gamma (s,\bar{q}_\tau ;\eta ,\tilde{q}_\tau )
\label{a22}
\end{equation}
where $\gamma $$(s,\bar{q}_\tau ;\eta ,\tilde{q}_\tau )=\frac 12\{\omega
\left( s,\bar{q}_\tau \right) \delta (\eta )-\omega \left( \eta ,\tilde{q}%
_\tau \right) \delta (s)\}$, $\omega \left( s,q\right) $ is
\[
\omega \left( s,q\right) =\frac 4{(2\pi h)^\nu h}\int dq^{\prime }V\left(
q-q^{\prime }\right) \sin \left( \frac{2sq^{\prime }}h\right) +F\left(
q\right) \frac{d\delta \left( s\right) }{ds}
\]
$\delta (s)$ is the Dirac delta function, $\{\bar{q}_\tau (\tau ;p_1,q_1,t),$
$\bar{p}_\tau (\tau ;p_1,q_1,t)\}$ and $\{\tilde{q}_\tau (\tau ;p_2,q_2,t),$
$\tilde{p}_\tau (\tau ;p_2,q_2,t)\}$ are pair of classical dynamical $pq$-
trajectories for 'positive' and 'negative' time direction and initial
condition taken at $\tau =t$:
\begin{eqnarray}
d\bar{p}/d\tau &=&\frac 12F(\bar{q}_\tau (\tau ));\bar{q}_t(t;p_1,q_1,t)=q_1
\nonumber \\
d\bar{q}/d\tau &=&\frac 1{2m}\bar{p}_\tau (\tau );\ \ \bar{p}%
_t(t;p_1,q_1,t)=p_1  \nonumber \\
d\tilde{p}/d\tau &=&-\frac 12F(\tilde{q}_\tau (\tau ));\tilde{q}%
_t(t;p_2,q_2,t)=q_2  \label{a25} \\
d\tilde{q}/d\tau &=&-\frac 1{2m}\tilde{p}_\tau (\tau );\tilde{p}%
_t(t;p_2,q_2,t)=p_2  \nonumber
\end{eqnarray}

The initial conditions $\bar{W}\left( p_1,q_1;p_2,q_2;E\right) $ are
connected with initial functions $\bar{W}\left( p_1,q_1;p_2,q_2;i\hbar \beta
\right) $ in analogous way:
\begin{eqnarray*}
\bar{W}\left( p_1,q_1;p_2,q_2;i\hbar \beta \right)  &=&\int dE\exp \left(
-\beta E\right) \bar{W}\left( p_1,q_1;p_2,q_2;E\right) = \\
&&\int dE\exp \left( -\beta E\right) \frac 1{2\pi }\int_{-\infty }^\infty
d\omega \exp \left( \frac{\beta \hbar \omega }2\right) k\left( \omega
,E\right)
\end{eqnarray*}
For checking the basic ideas of this approach we have used the classical
approximation of the initial condition of spectral density $\bar{W}\left(
p_1,q_1;p_2,q_2;E\right) $ in the following form:
\begin{equation}
\bar{W}\left( p_1,q_1;p_2,q_2;E\right) \approx \delta \left( E-H\left(
p_1,q_1\right) \right) \eta \left( -q_1^x\right) \delta \left(
p_1-p_2\right) \delta \left( q_1-q_2\right)   \label{mcini}
\end{equation}
where $H\left( p_1,q_1\right) $ is the classical Hamiltonian of the system.
The initial spectral density is non zero only in the half - space of
negative part of the x - axis.

Let us rewrite the integral equation (\ref{a22}) and the iteration form of
its solution in symbolic form: $W^t=\bar{W}^t+K_\tau ^tW^\tau $ and
\begin{equation}
W^t=\bar{W}^t+K_{\tau _1}^t\bar{W}^{\tau _1}+K_{\tau _2}^tK_{\tau _1}^{\tau
_2}\bar{W}^{\tau _1}+K_{\tau _3}^tK_{\tau _2}^{\tau _3}K_{\tau _1}^{\tau _2}%
\bar{W}^{\tau _1}+...  \label{r3}
\end{equation}
Here $\bar{W}^t$ and $\bar{W}^{\tau _1}$ is the quantum initial density
evolving classically in intervals $\left[ 0,t\right] $ and $\left[ 0,\tau
_1\right] $, while $K_{\tau _i}^{\tau _{i+1}}$ are operators, which describe
propagation between times $\tau _i$ and $\tau _{i+1}$. The time correlation
functions are the linear functionals of the spectral density:
\begin{equation}
\begin{array}{c}
C_{F\eta }\left( t,E\right) ==\frac 1{\left( 2\pi \hbar \right) ^{2\upsilon
}}\int \int dp_1dq_1dp_2dq_2\frac 12\left( F\left( p_1^x,q_1^x\right) \eta
\left( q_2^x\right) +F\left( p_2^x,q_2^x\right) \eta \left( q_1^x\right)
\right) \times  \\
W\left( p_{1,}q_1;p_{2,}q_2;t;E\right) = \\
\left( \phi |\bar{W}^t\right) +\left( \phi |K_{\tau _1}^t\bar{W}^{\tau
_1}\right) +\left( \phi |K_{\tau _2}^tK_{\tau _1}^{\tau _2}\bar{W}^{\tau
_1}\right) +\left( \phi |K_{\tau _3}^tK_{\tau _2}^{\tau _3}K_{\tau _1}^{\tau
_2}\bar{W}^{\tau _1}\right) +...
\end{array}
\label{a31}
\end{equation}
where $\phi \left( p_{1,}q_1;p_{2,}q_2\right) =\frac 12\left( F\left(
p_1^x,q_1^x\right) \eta \left( q_2^x\right) +F\left( p_2^x,q_2^x\right) \eta
\left( q_1^x\right) \right) $, brackets $\left( |\right) $ for functions and
$\bar{W}(\bar{p}_0,\bar{q}_0;\tilde{p}_0,\tilde{q}_0;E)$ or $K_{\tau
_i}^tK_{\tau _{i-1}}^{\tau _i}...K_{\tau _1}^{\tau _2}\bar{W}^{\tau _1}$
mean the integration over the phase spaces $\left\{
p_{1,}q_1;p_{2,}q_2\right\} $.

Introduced in the first part of this paper \cite{can1} the recurrent
relations for the pieces of dynamic trajectories according to (\ref{a25})
\begin{equation}
\begin{array}{c}
\bar{p}_{j-1}^j=\bar{p}(\tau _{j-1};\bar{p}_j^{j+1}-s_j,\bar{q}_j^{j+1},\tau
_j) \\
\bar{q}_{j-1}^j=\bar{q}(\tau _{j-1};\bar{p}_j^{j+1}-s_j,\bar{q}_j^{j+1},\tau
_j) \\
\tilde{p}_{j-1}^j=\tilde{p}(\tau _{j-1};\tilde{p}_j^{j+1}-\eta _j,\tilde{q}%
_j^{j+1},\tau _j) \\
\tilde{q}_{j-1}^j=\tilde{q}(\tau _{j-1};\tilde{p}_j^{j+1}-\eta _j,\tilde{q}%
_j^{j+1},\tau _j)
\end{array}
\label{a33}
\end{equation}
allow to obtain the explicit expression of the terms of series (\ref{r3})
and to analyze its mathematical structure. So for example the third term $%
\left( j=3\right) $ can be written as:
\[
K_{\tau _2}^tK_{\tau _1}^{\tau _2}\bar{W}^{\tau _1}=\int_0^td\tau
_2\int_0^{\tau _2}d\tau _1\int ds_2d\eta _2\int ds_1d\eta _1\times
\]
\[
\times \gamma (s_2,\bar{q}_2;\eta _2,\tilde{q}_2)\gamma (s_1,\bar{q}%
_1^2;\eta _1,\tilde{q}_1^2)\bar{W}(\bar{p}_0^1,\bar{q}_0^1;\tilde{p}_0^1,%
\tilde{q}_0^1;E)
\]

Note that average values of quantum operators $\bar{A}\left( t\right) $ can
be formally presented in the form analogous to (\ref{a31}) :
\[
\bar{A}\left( t,E\right) =\tilde{Z}^{-1}Tr\left( \exp \left( i\hat{H}t/\hbar
\right) \hat{A}\exp \left( -i\hat{H}t/\hbar \right) \delta \left( E-\hat{H}%
\right) \right) =
\]
\[
\frac 1{\left( 2\pi \hbar \right) ^{2\upsilon }}\int \int
dp_1dq_1dp_2dq_2\frac 12\left\{ A\left( p_1,q_1\right) +A\left(
p_2,q_2\right) \right\} \times
\]
\[
\times W\left( p_{1,}q_1;p_{2,}q_2;t;E\right)
\]

\section{Wigner approach in the theory of classical wave propagation in
random media}

In the case of the scale of inhomogeneities is large in comparison with the
wave length it is possible to neglect the backscattering and depolarization
of waves and to describe the classical wave propagation by the parabolic
wave equation \cite{krav} :
\begin{equation}
i\frac{\partial u}{\partial z}=-\frac 12\triangle u-\tilde{\epsilon}\left(
z,R\right) u;u\left( z_0,R,\omega \right) =u^0\left( R,\omega \right)
\label{schr1}
\end{equation}
where $k^{-1}$ is taken as a unit of length, $k=\omega \left( \bar{%
\varepsilon}\right) ^{\frac 12}/c$, $\omega $ is the wave frequency, $z$ is
the initial direction of wave propagation, $\triangle =\left( \frac \partial
{\partial x}\right) ^2+\left( \frac \partial {\partial y}\right) ^2$, $%
\tilde{\epsilon}\left( z,R\right) =\left( \varepsilon -\bar{\varepsilon}%
\right) /2\bar{\varepsilon}$, $\varepsilon $ is the dielectric permittivity,
the overbar denotes averaging over the medium fluctuations, $u\left(
z,R,\omega \right) $ is the slowly varying complex amplitude of the wave
field \cite{krav} , $u^o$ is the 'initial condition' for $u$ at $z=z_0$ and $%
R=\left( x,y\right) $ is a radius- vector in transverse plane to the initial
direction of the wave propagation (axis $z$ ). The Wigner- Liouville
distribution function is defined by

\[
\begin{array}{c}
f\left( z,P,Q,\omega \right) =\frac 1{\left( 2\pi h\right) ^\nu }\int \rho
\left( z,Q-\frac \xi 2,Q+\frac \xi 2,\omega \right) \exp \left( iP\xi
\right) d\xi ; \\
\rho \left( z,Q-\frac \xi 2,Q+\frac \xi 2,\omega \right) =u\left(
z,R_1,\omega \right) u^{*}\left( z,R_2,\omega \right) =u\left( z,Q-\frac \xi
2,\omega \right) u^{*}\left( z,Q+\frac \xi 2,\omega \right)
\end{array}
\]
Taking the time derivatives of the Wigner- Liouville distribution function
and using equation (\ref{schr1}) it is possible to obtain the integral
equation for $f\left( z,P,Q,\omega \right) $ in the form:

\[
\begin{array}{c}
f\left( z,P,Q,\omega \right) =f^0\left( z_0,\bar{P}_0,\bar{Q}_0,\omega
\right) +\int_0^zd\tau \int dSf\left( \tau ,\bar{P}_\tau -S,\bar{Q}_\tau
,\omega \right) \varpi \left( \tau ,S,\bar{Q}_\tau \right) , \\
\varpi \left( \tau ,S,Q\right) =-\frac 4{(2\pi h)^\upsilon h}\int dQ^{\prime
}\tilde{\epsilon}\left( \tau ,Q-Q^{\prime }\right) \sin \left( \frac{%
2SQ^{\prime }}h\right) +\breve{F}_\tau \left( Q\right) \frac{d\delta \left(
S\right) }{dS}
\end{array}
\]
where $\tau \in \left[ 0,z\right] $ . The dynamic trajectories $\left\{ \bar{%
Q}_\tau (\tau ;P,Q,z),\bar{P}_\tau (\tau ;P,Q,z)\right\} $ with 'initial
conditions' $\left\{ P,Q\right\} $ at $\tau =z$ are defined by the following
equations:

\begin{equation}
\begin{array}{c}
d\bar{P}/d\tau =\breve{F}_\tau (\bar{Q}_\tau (\tau ));\ \bar{Q}_z(z;P,Q,z)=Q
\\
d\bar{Q}/d\tau =\bar{P}_\tau (\tau )/m;\ \ \bar{P}_z(z;P,Q,z)=P \\
\breve{F}_\tau (Q)=\partial \left( \tilde{\epsilon}\left( \tau ,Q\right)
\right) /\partial Q
\end{array}
\label{dyn2}
\end{equation}

So in parabolic approximation the variable $z$ can be formally considered as
the ''time variable'' in the Schrodinger like equation and the developed
approach can be used for calculations of the average intensity, wave fields
moments, the scintillation index and correlation functions characterizing
the classical wave scattering and propagation in random media. We'll
restrict our interests with linear functionals of Wigner function,
representing the mentioned above values \cite{krav} :
\begin{equation}
\left( \phi |f^z\right) =\left( \phi |\tilde{f}^z\right) +\left( \phi
|K_{\tau _1}^z\tilde{f}^{\tau _1}\right) +\left( \phi |K_{\tau _2}^zK_{\tau
_1}^{\tau _2}\tilde{f}^{\tau _1}\right) +...  \label{s12}
\end{equation}
where brackets $\left( |\right) $ mean integration of functions $\phi $ and $%
f\left( z,P,Q,\omega \right) $ or $K_{\tau _j}^zK_{\tau _{j-1}}^{\tau
_j}...K_{\tau _1}^{\tau _2}\tilde{f}^{\tau _1}$ over phase space $\left(
P,Q\right) $. This represents also the averaged value of operators $\hat{\phi%
}\left( \hat{P},\hat{Q}\right) $ in the Wigner representation
\[
\bar{\phi}\left( z, \omega \right) =\frac 1{2\pi }\int dPdQ\phi \left( P,Q\right)
f\left( z,P,Q,\omega \right) =\left\langle u \left| \hat{\phi}\left( \hat{%
P},\hat{Q}\right) \right| u \right\rangle
\]
\begin{equation}
\phi \left( P,Q\right) =\int d\xi \exp \left( iP\xi /\hbar \right)
\left\langle Q-\frac \xi 2\left| \hat{\phi}\left( \hat{P},\hat{Q}\right)
\right| Q+\frac \xi 2\right\rangle   \label{s13}
\end{equation}
where $\left\langle Q-\frac \xi 2\left| \hat{\phi}\left( \hat{P},\hat{Q}%
\right) \right| Q+\frac \xi 2\right\rangle $ is a matrix element of operator
$\hat{\phi}$ \cite{tatr1} .

\section{Quantum and wave dynamics}

The possibility to convert series like (\ref{a31}) and (\ref{s12}) into the
form convenient for probabilistic interpretation allow us to develop the
Monte Carlo method for its calculation \cite{filmd1} , \cite{filmd2} .
Let us note only that the
ergodic hypothesis allow us to perform the integration on variables $\left\{
p_1,q_1;p_2,q_2\right\} $ for averaging in micro canonical ensemble
according to (\ref{mcini}), (\ref{a31}) by using the classic molecular
dynamics method. As has been shown in \cite{ulnb} the transition from the
integration over the phase- space at the fixed total energy (due to (\ref
{mcini})) to the time averaging along the classical dynamic trajectory
requires the correction factor $\left| \bigtriangledown H\left(
p_1,q_1\right) \right| $, which in our case should be added to the weight
function $\Omega $ of the $\rho $ - trajectories \cite{filmd1},
\cite{filmd2}.
Here multidimensional vector $\bigtriangledown
H\left( p_1,q_1\right) $ has the following components
\[
\begin{array}{c}
\{\partial H/\partial p_1^1,...,\partial H/\partial p_1^\upsilon ,\partial
H/\partial q_1^1,...,\partial H/\partial q_1^\upsilon ,..., \\
\partial H/\partial p_N^1,...,\partial H/\partial p_N^\upsilon ,\partial
H/\partial q_N^1,...,\partial H/\partial q_N^\upsilon \}
\end{array}
\]
for the $\upsilon $ - dimensional space, while the dynamic evolution has
been realized according to equations (\ref{a33}).

\section{Numerical results}

\subsection{Electron spectra}

The Fig. \ref{fig:crqd1} shows the time correlation function $C_{F\eta
}\left( t,E\right) $ for $E/V_0=-.92$. From physical point of view this
function characterizes the momentum flow of quantum particles tunneling
through the potential barrier. First of all it is necessary to note that at
small (less than 20 $t\hbar /V_0$) and large times (more than 20 $t\hbar
/V_0 $) the time correlation function has two different frequencies of
oscillation. Physically this behaviour can be explained by the shape of our
double well, which have two space scales and two energy scales. The left and
right shallow narrow wells separated by barrier have the space scale of the
order $\sigma $ and the depth equal to $0.12*V_0$. The space scale of the
large well containing both shallow narrow wells and the barrier is $\tilde{%
\sigma}$, while the depth is $V_0$. These scales satisfies as we mentioned
before to the following inequalities: $0.12=V_0^{\prime }/V_0\ll 1$ and $%
\tilde{\sigma}>\sigma $.

The main contribution to the time correlation function at initial time (less
20) comes from the quantum trajectories with virtual energy close to the
height of the barrier, when they path trough the top of the barrier. Due to
the momentum jumps these trajectories can penetrate in both shallow wells
and can be also trapped there. So the high frequency oscillations can be
connected with multiple reflections and tunneling transitions of these
trajectories . It is necessary to stress that derivative of this correlation
function with respect to time is the momentum- momentum time correlation
function taken at the top of the barrier. So from physical point of view
these oscillation results from changing of the direction of the main
momentum flow of quantum particle at multiple tunneling transitions.

The same functions obtained in approximation of the classical trajectories
(the first term of the iteration series) are identically equal to zero.
Indeed at the averaging in the micro- canonical ensemble the initial $p,q$
data for our classical and quantum trajectories have been taken from the
classical trajectory moving in the left part of our double well. That means
that the initial energy of our trajectories was lower than the height of the
barrier and the classical trajectories without momentum jumps were unable to
leave the left part of our double well and consequently to give contribution
to the considered time correlation function.

At the large time (more than 20 $t\hbar /V_0$) the main contribution to the
time correlation function considered comes from the trajectories with the
virtual energy far (much more) from the height of the barrier. So these
trajectories can move anywhere in the large well and that is the reason of
low frequency oscillations of the time correlation function.

The next Fig. \ref{fig:spect2} and Fig. \ref{fig:spect} present the squared
amplitude of the Fourier transform of the time correlation function $%
|k\left( \omega ,E\right) |^2$ versus the frequency $h\omega /V_0$.
Analytical semi classical estimations of the positions of the sharp peaks of
the function $k\left( \omega ,E\right) $ are presented by stars 1.
Calculated positions of the sharp peaks of the $|k\left( \omega ,E\right)
|^2 $ on the Fig. \ref{fig:spect2} and Fig. \ref{fig:spect} are in a
agreement with analytical estimations.


\subsection{Position and momentum dispersions at quantum tunnelling}

The next Fig. \ref{fig:dxcld1} and Fig. \ref{fig:dxqd1} present results for
position dispersions obtained for classical and quantum trajectories
respectively versus dimensionless time $tV_0/\hbar $ for 1D case. As we
mentioned before the initial energy of the our trajectories was lower than
the height of the barrier and calculations of position dispersion allowing
for only the first term of the iteration series were implemented by making
use of only the classical trajectories, which were unable to leave the left
shallow well. These results for position dispersion are presented on the
Fig. \ref{fig:dxcld1}. The squared root of the position dispersion gives the
estimation of the oscillation amplitude (of order of 0.05$k^{-1}$) of these
trajectories and allow to estimate the characteristic size of the available
space in the left potential well for these trajectories. Note that the
classical dynamics gives non damping oscillations of position dispersion.

The quantum trajectories give the qualitatively different behaviour of
position dispersion. These results are presented on Fig. \ref{fig:dxqd1}.
The virtual energy of quantum trajectories can be larger than the height of
the barrier, so these trajectories can move in the right part of the double
well, where the virtual energy of these trajectories can become smaller than
the height of the barrier and these trajectories can be trapped there.
At any case the quantum trajectories can travel anywhere in the double well
and even leave it. So the position dispersion of quantum trajectories is
much larger than the same value of classical trajectories. However position
dispersion of the quantum trajectories has a very interesting peculiarity.
At time $tV_0/\hbar =30$ the position dispersion becomes practically equal
to zero. It is interesting that this happens at time when time correlation
function changes the frequency of its oscillations. Note that at this time
the momentum dispersion presented on next Fig. \ref{fig:dpd1} (curve 2)
changes also its characteristic behaviour after sharp oscillation. One can
see that the minima of position and momentum dispersions happens
approximately at the same time. Then the momentum dispersion practically
stabilizes.

Our physical explanation of this peculiarity is connected not only with
tunneling transitions of quantum particles but also with existing of a quasi
stationary state at the top of the narrow barrier.
This state may be the reason that both position and momentum dispersions at
the same time (of about 30 $tV_0/\hbar $ ) have minima. The time of life of
this quasi stationary state may be estimated from the Fig. \ref{fig:dxqd1}
as of order about 5$tV_0/\hbar $


The next Fig. \ref{fig:dpd1} presents the momentum dispersion for classical
and quantum trajectories. The momentum dispersion for classical trajectories
oscillates with very small amplitude (curve 1).

\section{Classical wave propagation in 2D and 3D waveguides}

To test the developed stochastic dynamics approach the wave propagation
along z-axis $\left( z\geq 0\right) $ in 2D and 3D waveguides with realistic
profile of refractive index has been investigated in parabolic
approximation. Gaussian beam distribution was used as a initial condition at
$z=0$. Figures \ref{fig:Fig10} and \ref{fig:Fig11} present the 3D case
numerical data for average position $\left( \bar{x}\left( z\right), \bar{y}%
\left( z\right) \right) $ and dispersion $\beta =\bar{R\left( z\right)^2}-%
\bar{R}\left(z\right) ^2$ of Gaussian beam vs distance $z$. Foci points
along waveguide are indicated by the minimum values of $\beta $.

\section{Conclusion}

In the Wigner formulation of quantum statistical mechanics for canonical and
micro canonical ensembles we have presented a new computational technique
allowing quantum dynamics simulations for systems including subsystems of
quantum interacting particles and subsystems of classical heavy scatterers
as well as the system of quantum particles in external potential field. The
developed approach for quantum dynamics includes a sophisticated combination
of well known molecular dynamics method and Monte Carlo technique.
Numerical results have been presented for the time correlation function
characterizing the energy level structure, the momentum flow of tunneling
particles at barrier crossing and the absorption spectra of electron in
potential well have been calculated.
Tunneling transitions and the effect of the quasi stationary
state has been considered as the reason of the peculiarities in behaviour of
the time correlation functions and position and momentum dispersions.

Possibility of applying the developed approach to the theory of classical
wave propagation in random media have been also considered. For classical
waves some results have been obtained for Gaussian beam propagation in 2D
and 3D waveguides.

\section{Acknowledgments}

The authors is very appreciated to Professor K. Singer for fruitful
discussions, invaluable comments and interest to work. The authors expresses
thanks to Russian Fund for Basic Researches for financial support of this
work ( grants 97- 02- 16572, 97-1-00931, 96-1596462 ).


\newpage
\begin{figure}
\caption[Momentum]
       { \label{fig:crqd1}
The time correlation function $C_{F \eta}\left( t,E\right)$ versus
$tV_0/ \hbar$ for quantum trajectories.
($ E /V_0 = -0.92 $)}
\caption[Momentum]
       { \label{fig:spect2}
$|k\left( \omega ,E\right)|^2$ vz frequency $h\omega /V_0$:
1- analitical estimations of the sharp peak positions;
2 - numerical results;
3- the error bar; $\sigma=0.26$, $\tilde \sigma= 2.23$.}
\end{figure}

\begin{figure}
\caption[Momentum]
   { \label{fig:spect}
$|k\left( \omega ,E\right)|^2$ vz frequency $h\omega /V_0$:
1- analitical estimations of the sharp peak positions;
2 - numerical results;
3- the error bar; $\sigma=0.26$, $\tilde \sigma= 3.43$.}
\caption[posdcl]
   { \label{fig:dxcld1}
Position dispersion: 1- classical trajectories;
($ E /V_0 = -0.92 $)}
\end{figure}

\begin{figure}
\caption[posdq]
   { \label{fig:dxqd1}
Position dispersion: 1- quantum trajectories;
($ E /V_0 = -0.92 $)}
\caption[Momentum]
   { \label{fig:dpd1}
Momentum dispersion: 1- classical trajectories; 2- quantum trajectories;
($ E /V_0 = -0.92 $)}
\end{figure}

\begin{figure}
\caption[waveguide]
       { \label{fig:Fig10}
Average position $\bar{x}(z)$ and $\bar{y}(z)$ vs z for 3D waveguide.
Starting point
at $z=0$ corresponds approximately to point (21,21), 'center' of waveguide
is nearly the point (17,17).}
\end{figure}

\begin{figure}
\caption[dispqp]
       { \label{fig:Fig11}
$\beta =\bar{R^2 \left( z\right)} -\left( \bar{R}\left(z\right) \right) ^2$
vs z (in discrete units ) for 3D waveguide ($R$ =(x,y)). }
\end{figure}






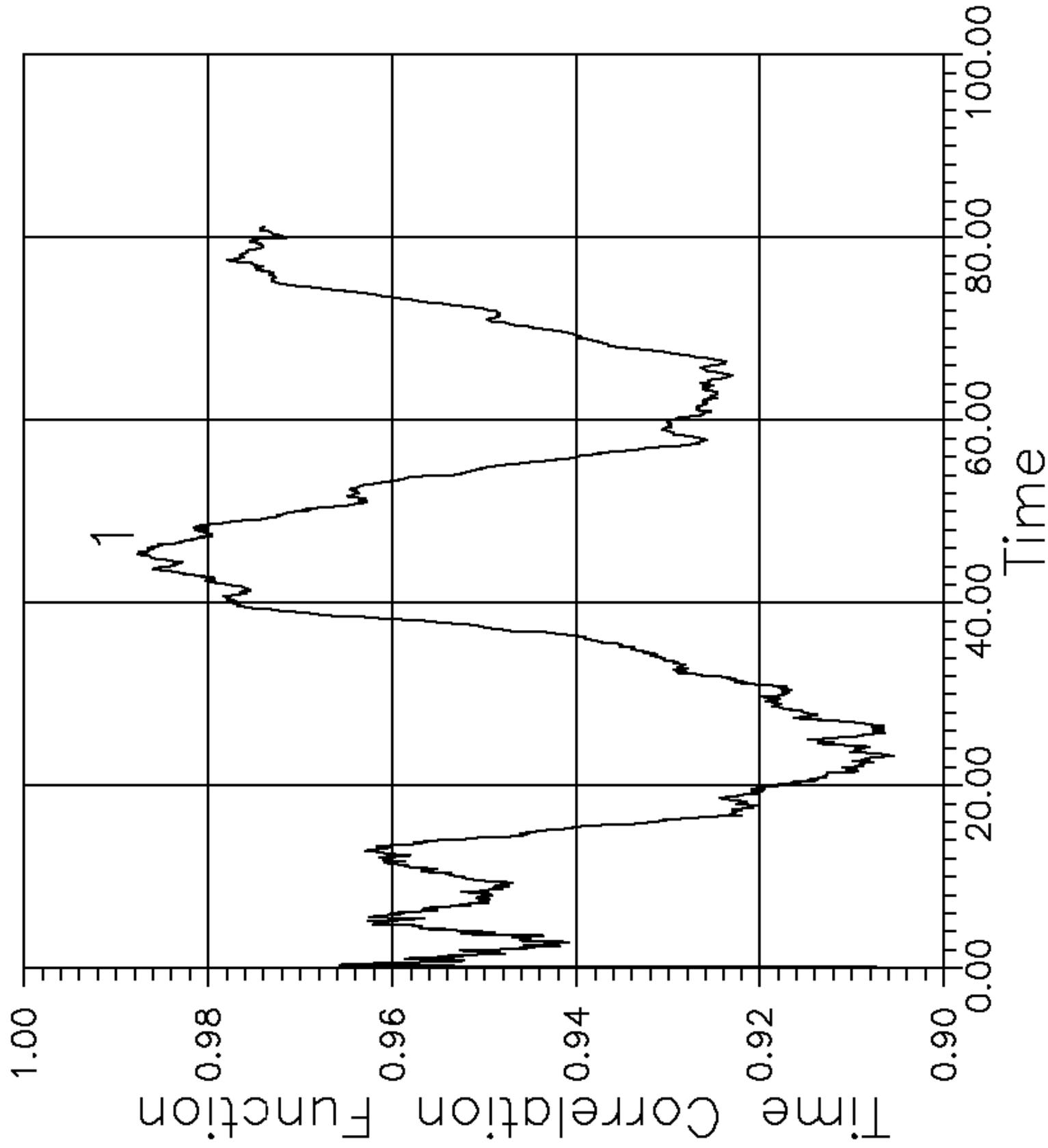
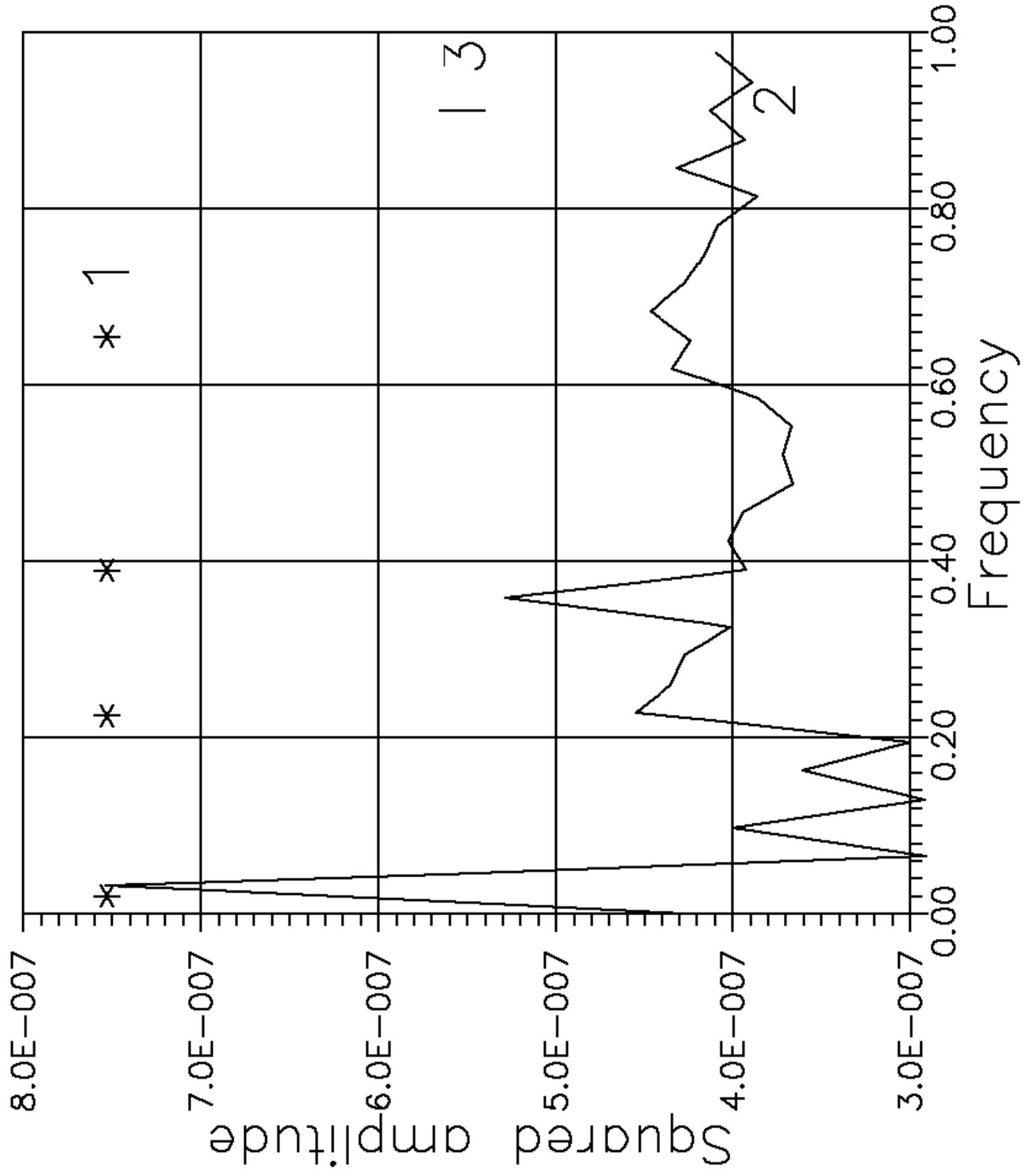
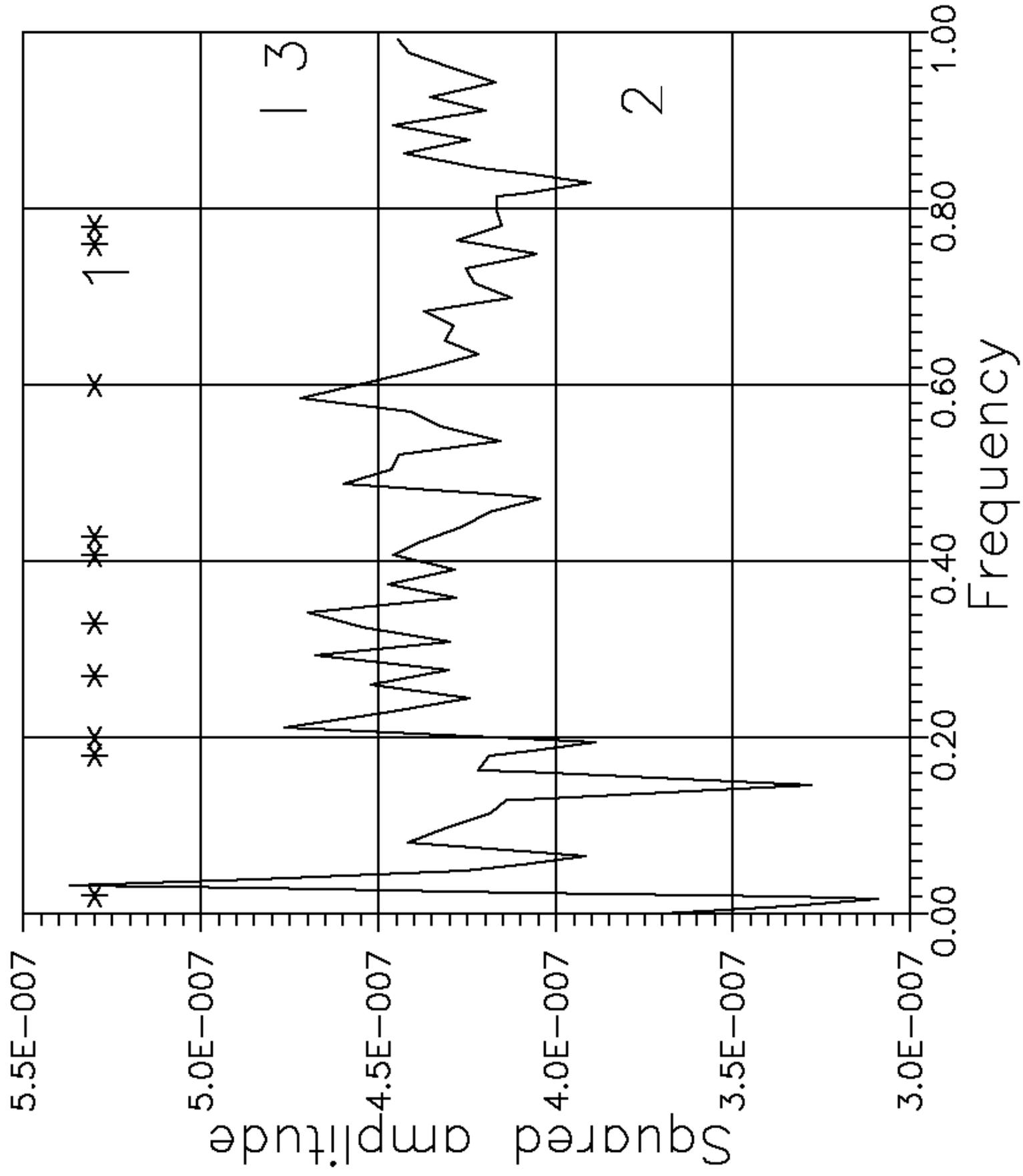
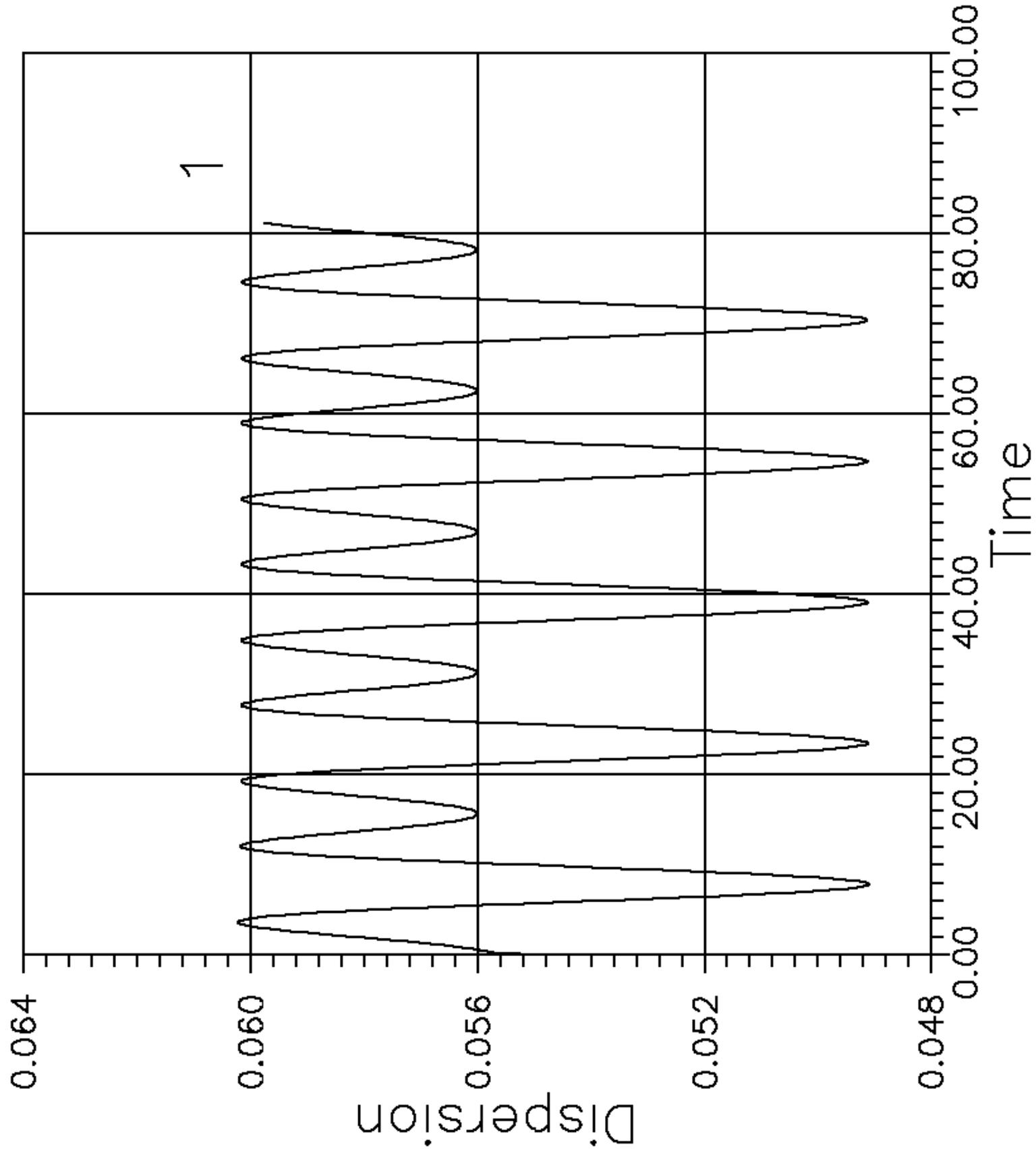
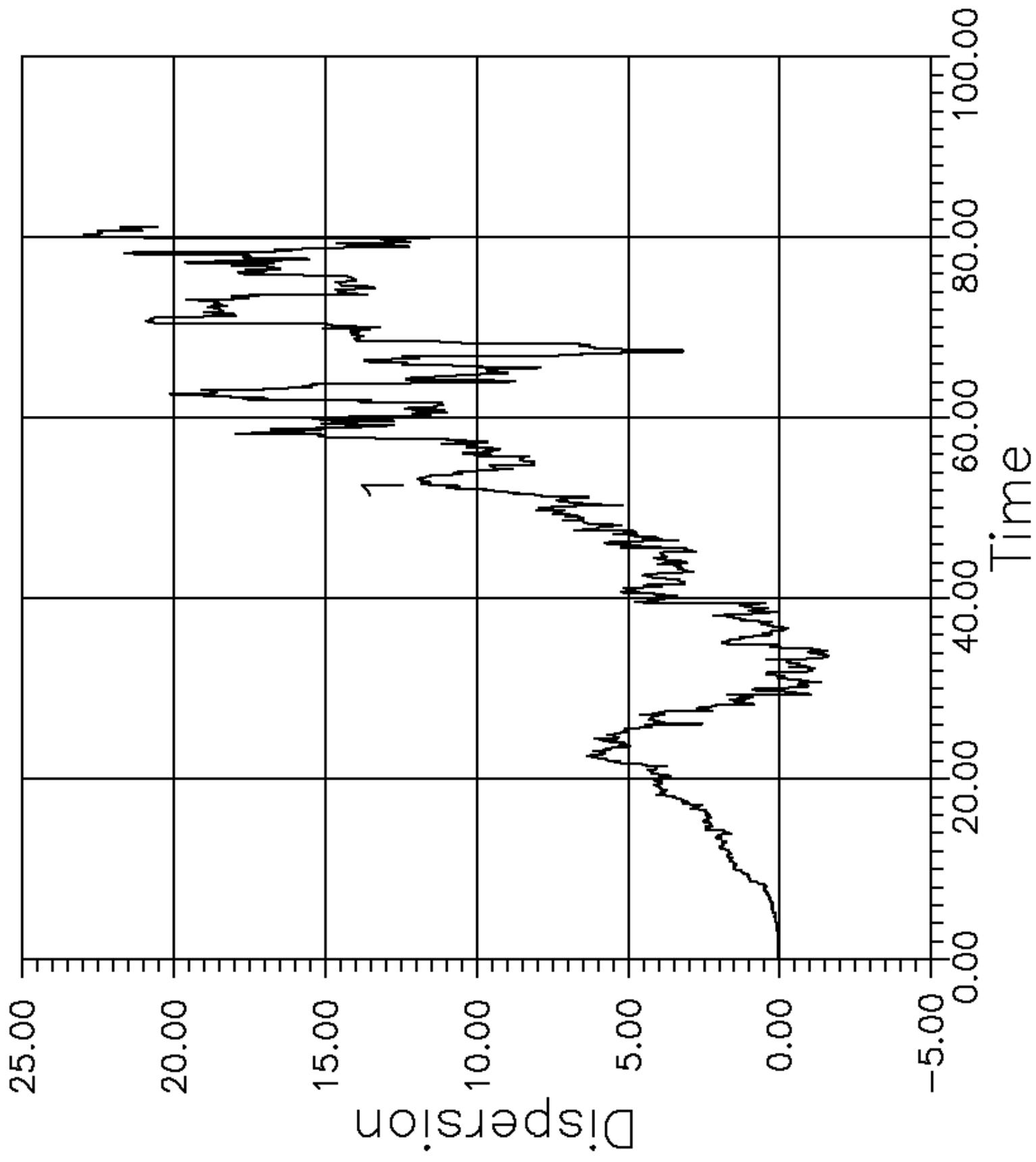
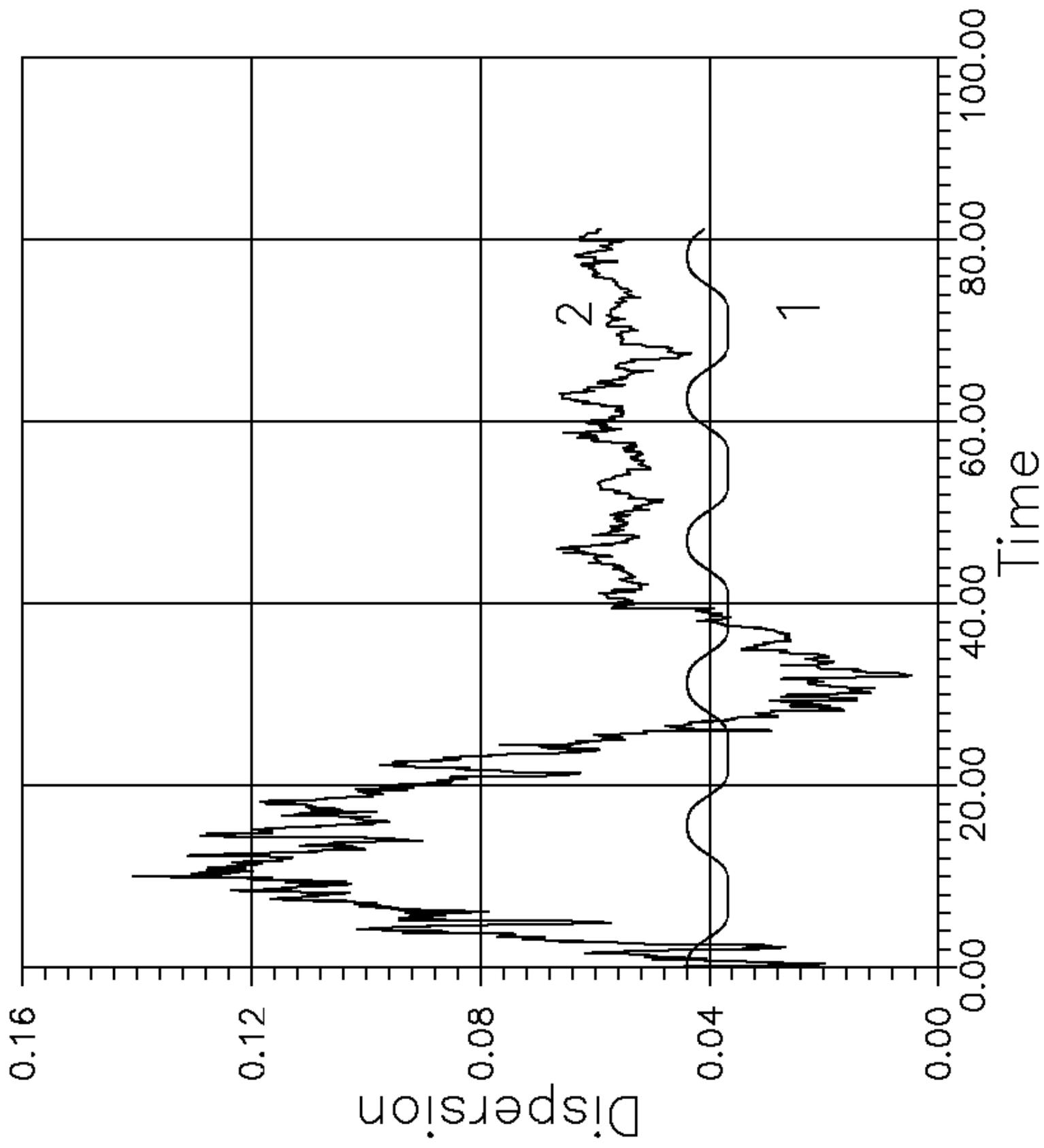
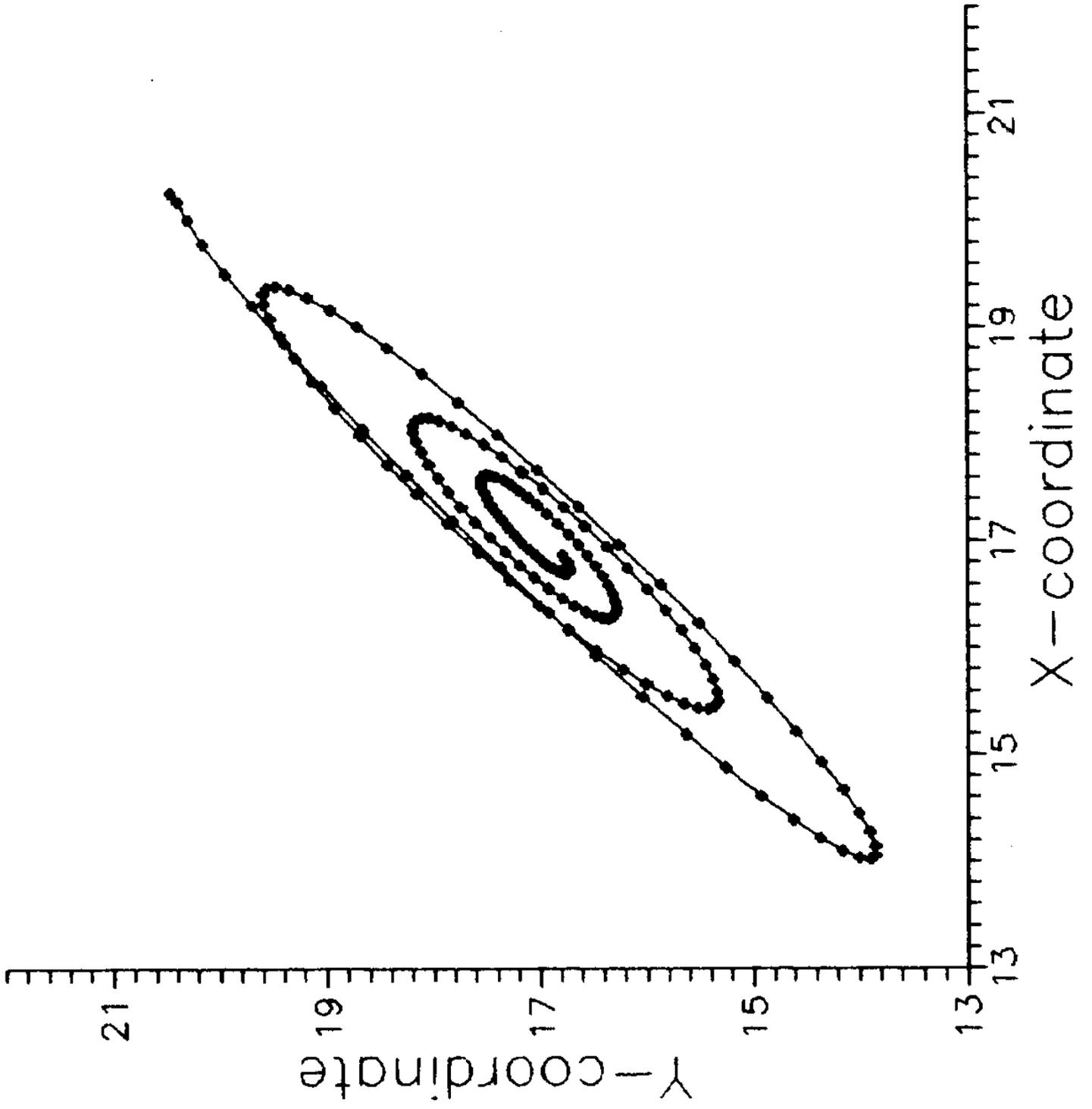
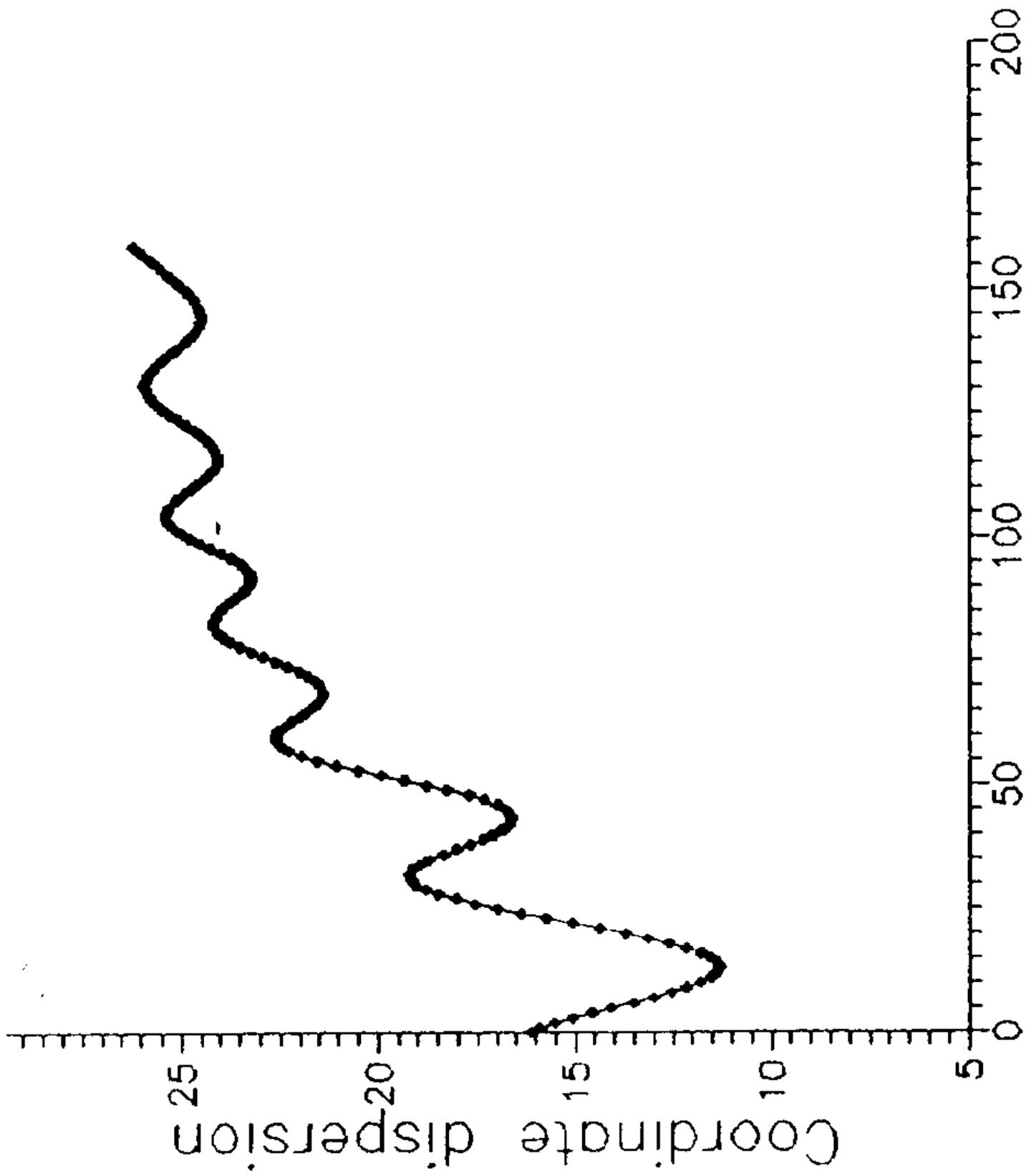
\end{document}